\documentclass[12pt]{article}
\usepackage{geometry}
\DeclareMathAlphabet{\mathpzc}{OT1}{pzc}{m}{it}
\usepackage{amssymb}
\usepackage{epstopdf}
\usepackage{amsmath}
\usepackage{graphicx}
\usepackage{stmaryrd}
\usepackage{hyperref}
\usepackage[normalem]{ulem}
\usepackage{url}

\newcommand{\iden}{{\bf 1}}
\newcommand{\ket}[1]{\vert #1\rangle}
\newcommand{\bra}[1]{\langle #1\vert}
\newcommand{\be}{\begin{equation}}
\newcommand{\ee}{\end{equation}}

\title{A Proof of Specker's Principle}
\author{Guido Bacciagaluppi\footnote{Descartes Centre for the History and Philosophy of the Sciences and the Humanities and Freudenthal Institute, Utrecht University; IHPST and SPHERE, Paris; email: g.bacciagaluppi@uu.nl\ .} }

\date{16 November 2023}

\begin{document}
\maketitle

\begin{abstract}
Specker's principle, the condition that pairwise orthogonal propositions must be jointly orthogonal (or rather, the `exclusivity principle' that follows from it), has been much investigated recently within the programme of finding physical principles to characterise quantum mechanics. Specker's principle, however, largely appears to lack a physical justification. In this paper, I present a proof of Specker's principle from three assumptions (made suitably precise): the existence of maximal entanglement, the existence of non-maximal measurements, and no-signalling. I discuss these three assumptions and describe canonical examples of non-Specker sets of propositions satisfying any two of them. These examples display analogies with various approaches to the interpretation of quantum mechanics, including retrocausation. I also discuss connections with the work of Popescu and Rohrlich. The core of the proof (and the main example violating no-signalling) is illustrated by a variant of Specker's tale of the seer of Nineveh, with which I open the paper.

{\bf Keywords}: Specker's principle, reconstructions of quantum mechanics, retrocausation, Popescu-Rohrlich boxes
\end{abstract}

\newpage

\section*{Specker's seer meets Popescu and Rohrlich}\label{SPR}
At the Assyrian school for prophets in Arba'ilu, there taught in the age of king Asarhaddon a sage from Nineveh. He was an outstanding representative of his discipline (solar and lunar eclipses), who except for the heavenly bodies had thoughts almost only for his \emph{two} daughters. His teaching success was modest; the discipline was considered dry, and also required previous mathematical knowledge that was scarcely available. If thus in his teaching he failed to capture the attention he wanted from the students, he received it overabundantly in a different area: no sooner had his daughters reached the marriageable age than he was flooded with requests for their hands from students and young graduates. And even though he did not imagine wishing to keep them with him forever, yet they were still far too young, and the suitors in no way worthy of them. And so that they all should convince themselves of their unworthiness, he promised their hands to those two who could perform a set prophecy task. The suitors were led in front of two tables on each of which stood three boxes in a row, and urged to say which boxes contained a gem and which were empty. Yet, as many as would attempt it, it appeared impossible to perform the task. Indeed, after they both had made their prophecies, the father urged each of the suitors to open two boxes that they had named as both empty or as both not empty: it always proved to be that one contained a gem and the other did not, and in fact the gem lay now in the first, now in the second of the boxes that were opened; and yet, every time the suitors both opened the first box, or both the second, or both the third, whatever one of them  found (or failed to find) in one box, the other would also find (or fail to find) in the corresponding box -- which showed that the gems needs must have been contained in the boxes in the first place. But, then, how should it be possible, out of three boxes, to name no two as empty or as not empty? Thus indeed the daughters would have remained unmarried until their father's death, had they not encouraged two bright young students with whom they were secretly in love and whose names, let it be known, were Sandu and Daniel, to attempt the task. Now, Sandu and Daniel were not renowned at the time as having any particular gift for prophecy, but they were very ingenious and hard-working, and Daniel's uncle was a prophet of considerable standing, so Daniel hoped to have inherited some of the same gift. Besides, the two friends were both desperately in love: so they picked up their courage and sought an audience with their beloved ones' father. When the father heard Sandu and Daniel asking him for his daughters' hands, he smiled to himself, and (judging they had not the slightest chance of succeeding in the task) declared that not only were they welcome to come upon the morrow to test their skills, but that he would give them not one, nor two, but one hundred chances if they should care to try. And so the father was awake all night setting up two hundred little tables and six hundred little boxes and placing gems from his collection in at least two hundred of them (for he had won many prizes in prophecy competitions over the years). And the next morning Sandu and Daniel duly presented themselves at the sage's home and embarked on making prophecies. As soon as they had both performed a set of prophecies, at the father's urging each of them opened two boxes that they had named as both empty or as both not empty, and, lo and behold: it always proved to be that one contained a gem and the other did not, and the gem lay now in the first, now in the second of the boxes that were opened; and yet, every time they both opened the first box, or both the second, or both the third, whatever one of them found (or failed to find) in a box, the other would also find (or fail to find) in the corresponding box -- just as had happened to every pair of suitors before them (and as the two daughters had long grown accustomed to). And so they laboured all morning (for prophesying is a strenuous task, especially if you are still a student), becoming ever more disconsolate at each unsuccessful attempt. At the end of fifty trials, the father called a break, and kindly offered Sandu and Daniel some tea and cucumber sandwiches of which they partook, grateful of a rest and of the opportunity to exchange some precious words with their beloved ones. When the hour was over and the final fifty trials were to begin, the younger of the two daughters walked up to her father and spoke to him thus: `Father, Sandu and Daniel realise that you are a far better prophet than they are, and that if it is you who choose the boxes that are to be opened, then in no single trial will their prophecies stand. My sister and I therefore humbly beseech you (for we are very fond of them, even though you must be laughing at their youth and inexperience) to give them one small chance of success. If you will, demand they be successful in all fifty trials that still lie before them, but please let them choose themselves which boxes shall be opened and prophesy which shall be found empty or full. If they both fail, they will lay down any claim to our hands forever. But if at least one of them succeeds (for neither of us could ever be happy if the other is not), oh! please let your loving daughters be married to them, and we shall forever and a day be grateful to you, dear Father'. Now, well knowing how tiny the chance was of either Sandu or Daniel successfully predicting fifty trials at equal odds (unless they had a very great gift for prophecy, indeed), but equally moved by this eloquence (because the younger daughter who had thus spoken had her own subtle gift of winding her father around her little finger), the father consented to this alteration of the remaining trials, of course as long as Sandu and Daniel still made their prophecies before seeing the other's boxes opened. And so Sandu and Daniel embarked again on making prophecies for the rest of the day. Upon each trial, first Daniel announced which two boxes he would open (which by then all expected would contain one gem between them) and prophesied which one would be full and which one would be empty. Then Sandu chose his first box, prophesied whether this would be full or empty, and opened this box to verify whether or not his prophecy was correct. Then he chose a second box, which of course he correctly prophesied to be full if the first had been empty and empty if the first had been full. Finally Daniel opened his two chosen boxes and verified his earlier prophecy. Upon the first trial, both Sandu's and Daniel's prophecies proved to be successful, and so did they upon the second trial, and upon the third. Upon the fourth trial, however, Sandu's first prophecy was falsified, and in the subsequent trials it became apparent that the first of Sandu's prophecies met now with success, now with failure, as if he had no gift for prophecy at all (of course his second prophecy was always successful). But the father had already turned pale by the end of the second trial, realising he had been outsmarted. And, indeed, Daniel -- whether his family gift had found his way to him at last or through some other artifice, and to the increasing surprise and delight of the numerous bystanders (for by this time the father's servants and neighbours had long started gathering around) -- kept meeting with success, trial upon trial upon trial. Just before the hundredth and last trial, the father interjected, protesting weakly that he had not meant Sandu to announce his second prophecy only after having already verified the first one. To this now the elder of his daughters replied: what difference did that make to Sandu's chances of success, for it made no difference to his first prophecy, and the second stood or fell with the first? And so, grumbling, the father let the final trial proceed, and from the last of his boxes (which he had  prophesied to be full) Daniel triumphantly extracted a sparkling emerald that matched one that Sandu had just extracted from his. The four young people were married the very next day, and henceforth and for the rest of their lives both Sandu and Daniel enjoyed a reputation as formidable prophets. Meanwhile, the father consoled himself in the knowledge of having brought up two very clever daughters, indeed.

\section{Introduction}\label{intro}
`Specker's principle' can refer to a number of related conditions on sets of `propositions which are not simultaneously decidable' (Specker 1960). The original version of the condition is that sets of pairwise compatible propositions must be jointly compatible, and as such it enters as an axiom in the definition of partial Boolean algebras by Kochen and Specker (1965).\footnote{A(n infinitary) version of Specker's principle is also one of Mackey's (1963) axioms.} In the literature up to the 1980s this condition is usually known as `coherence',\footnote{Which, as pointed out by an anonymous referee, can be derived in turn from a condition called `regularity' (which itself follows easily if there are enough states to distinguish orthogonal propositions) together with closure of measurements under coarse-graining (Wilce 2009, Lemma 95 and Theorem 117).} and a basic result of quantum logic is that every coherent orthomodular poset is canonically isomorphic to a transitive partial Boolean algebra (Finch 1969, Gudder 1972). The version of the principle discussed in this paper is the strictly weaker one that propositions that are pairwise orthogonal (i.e.\ compatible and mutually exclusive) must be jointly orthogonal. It is known as `orthocoherence' in the earlier literature, and is a key feature of the discussion of tensor products in quantum logic (Foulis and Randall 1981).\footnote{For a helpful overview of classic issues in quantum logic, see Hughes (1989, Chapter 7).} An even weaker version of the principle, that pairwise orthogonal propositions have probabilities whose sum is bounded by 1, has been the object of recent discussion in the context of the search for physical principles characterising quantum mechanics, for instance in Cabello, Severini and Winter (2014), Fritz \emph{et al.}\ (2013), Henson (2012), and Cabello (2012), where it is referred to as the `exclusivity principle', `local orthogonality', `consistent exclusivity', and `Specker's principle', respectively. This version has proved extremely powerful in applications.\footnote{Perhaps especially in the hands of Ad\'{a}n Cabello: see Cabello (2013a, 2013b, 2015) and most strikingly Cabello (2019).} I shall follow most of the recent literature in referring to it as the `exclusivity principle' (or `E-principle'), and shall use `Specker's principle' and `strong Specker principle' to refer to the stronger conditions of  orthocoherence and coherence, respectively.

It is easy to construct toy examples that violate Specker's principle, and the prime one is Specker's (1960) own fable of the seer of Nineveh, adapted above and discussed in detail by Liang, Spekkens and Wiseman (2011). (A related example is the `firefly box', to which I return in Section~\ref{b}.) In the original variant, Specker tells the tale of a seer who sets a prophecy task to his (single) daughter's suitors. The suitors are asked to prophesy which of three boxes will reveal a gem when opened and which will be empty. After their prophecy, the father requires them to open two boxes that they had indicated as being both full or both empty, but of these one always turns out to be full and one empty, so that the prophecy is always falsified. One day, however, the daughter without warning opens two boxes predicted one to be full and one to be empty -- which turns out to be correct, while the third box cannot be opened.\footnote{There are two full translations of Specker's (1960) paper containing the original fable, by Stairs (1975) and by Seevinck (2011). Liang, Spekkens and Wiseman (2011) include an amalgam of the two. I have adapted my own translation of the fable from Bacciagaluppi (2016).} This is an example of a violation of Specker's principle (and the strong Specker principle), because one can always open boxes $A$ and $B$ together, or $B$ and $C$, or $C$ and $A$, and the results are always mutually exclusive, i.e.\ the corresponding propositions are pairwise orthogonal (and thus in particular pairwise compatible); but the three propositions are clearly not jointly compatible (and thus in particular not jointly orthogonal), because there is no joint probability distribution over the three corresponding events that would return the flat distribution $p=1/2$ for the three marginals.

While Specker's principle is easy to formulate, its physical justification has remained rather obscure. For instance, referring to the strong Specker principle, Kochen himself has stated  (as reported in Cabello 2012): `Ernst and I spent many hours discussing the principle [...]. The difficulty lies in trying to justify it on general physical grounds, without already assuming the Hilbert space formalism of quantum mechanics. We decided to incorporate the principle as an axiom in our definition of partial Boolean algebras [...]. I have never found a general physical justification for [it]'.

A number of proofs of the exclusivity principle have been found in recent years (Chiribella and Yuan 2014; Barnum, M\"uller and Ududec 2014; Cabello 2019, Appendix~A; Chiribella \emph{et al.} 2020), but this paper provides a proof of Specker's principle in the sense used here of orthocoherence -- in fact a surprisingly simple one, deriving it from a combination of the existence of an analogue of maximally entangled states, the existence of an analogue of `non-maximal' measurements, and no-signalling. (These conditions will be made suitably precise below.)\footnote{There is also a proof of Specker's principle in H\"{o}hn (2017, Theorem 3.2) from what would appear to be stronger assumptions. I shall not attempt a full comparison with the extant proofs of exclusivity, but certain similarities ought to be pointed out. In particular, assumption (b) below (the existence of `non-maximal' measurements) seems to be quite close to the assumption of the existence of `ideal' or `sharp'  measurements in the proof by Chiribella and Yuan (2014), because as emphasised by Ad\'{a}n Cabello (2019, Appendix A) they assume that there are observables such that they and \emph{all their coarse-grainings} can be measured without disturbing any compatible observables (my thanks to Ad\'{a}n for pointing this out to me).} 

The derivation presented here can be seen as a variant of Popescu and Rohrlich's (1994) strategy of trying to derive quantum theory from a suitable combination of nonlocality and no-signalling, and there are further connections with Popescu and Rohrlich's work -- specifically with Popescu--Rohrlich nonlocal boxes (PR boxes), on which I shall also comment.

I present the main result in Section~\ref{proof}. Then, in Section~\ref{comments}, I discuss the three assumptions in turn, and give toy examples of failures of each of the three. These examples display analogies with various approaches to the foundations of quantum mechanics: respectively, collapse, pilot-wave theory, and retrocausal models. Finally, in Section~\ref{PR}, I briefly address the connection with PR boxes.

2020 marked the centenary of Ernst Specker's birth. I had the honour and the pleasure of having him as one of my most inspiring teachers at ETH Z\"{u}rich, as well as hearing the tale of the seer of Nineveh from himself sometime around the spring of 1985. I fondly and respectfully dedicate this paper to Professor Specker's memory.

\section{The proof}\label{proof}
The formulation of Specker's principle that we shall use is: for any $n$, if a set of propositions $A_1,\ldots,A_n$ are pairwise orthogonal, they are jointly orthogonal. 

As the framework for the proof we shall assume any propositional structure allowing for a suitable notion of orthogonality, in the sense that orthogonality of a pair or a set of propositions implies that the propositions are jointly compatible (i.e.\ there is an experiment in which they can be jointly measured) and mutually exclusive (i.e.\ in all experiments in which they can be jointly measured they are mutually exclusive in all states). We assume additionally that there is some appropriate notion of composition defined for it, which we symbolically denote by a tensor product (as well as some suitable notion of systems being dynamically isolated from each other). 

While the exact details of this propositional structure do not matter,\footnote{For definiteness we could take orthoalgebras -- one of the best current candidates for providing an abstract setting for a generalised probability theory (for a good discussion, see Hardegree and Frazer (1981)). Every orthocoherent orthoalgebra is automatically an orthomodular poset (so every coherent orthoalgebra is automatically a transitive partial Boolean algebra) and if enough states exist one can define tensor products of orthoalgebras, even though such constructions need not preserve orthocoherence. Note that it is possible that more general structures such as effect algebras (Foulis and Bennett 1994) are incompatible with the assumptions of the proof, specifically assumption (b) -- which in fact is satisfied by the lattice of projections but not the poset of effects in Hilbert space (see Section~\ref{b}).} the proof does rely on there being a rich enough structure of \emph{experiments}. Specifically, we shall require that in the case of tensor products there be experiments that can be carried out via procedures only involving the relevant subsystem (i.e.\ such that one system remains isolated from the other subsystem), and that the probabilities for performing \emph{finite sequences of experiments} in any order (or simultaneously where possible) be well-defined.\footnote{In quantum mechanics, propositional systems may be given by projections or effects on Hilbert space, and projection-valued or effect-valued measures (resolutions of the identity) represent observables, i.e.\ exhaustive sets of jointly compatible propositions. Experiments are described by appropriate families of Kraus operators, which are closed under sequential composition (and for bipartite experiments can be implemented using interactions with the relevant subsystem but no interaction between the subsystems). The toy models of Section~\ref{comments} will also have enough structure to define probabilities for results of sequential measurements (or can be straightforwardly extended to this effect) -- even though not in a formally satisfactory way due to the lack of suitable `conditional states' (cf.\ footnote~17
).}
 
The more specific assumptions to be used in our proof are: 

(a) \emph{Existence of `maximal entanglement'}: We require that there exist a state on the composition of two copies of the same system in which any non-trivial proposition $A$ of a single system will have non-trivial probability,\footnote{That is, excepting any trivially true or trivially false propositions, which have probability 1 or 0 in \emph{all} states.} and any measurement results for copy 1 of the system and for copy 2 of the system will be perfectly correlated (even though the propositions of a single system need not be jointly compatible).

(b) \emph{Existence of `non-maximal measurements'}: For any proposition $A$, there is a set ${\cal C}_A$ of measurements of $A$ involving only the system in question, such that if the propositions $A_i$ are jointly compatible, performing any measurements in the respective ${\cal C}_{A_i}$ in any order (or simultaneously) yields a joint measurement of the propositions $A_i$.

(c) \emph{No-signalling}: No experimental procedure involving only one system may affect the probabilities for results of experimental procedures involving only another system if the two systems are isolated from each other.

We shall discuss these assumptions in detail in Section~\ref{comments}. Their conjunction will now enforce Specker's principle on the subsystems of the composite. 

Assume we have a system violating Specker's principle, i.e.\ a system with a set of $n$ pairwise orthogonal propositions $A_1\ldots,A_n$ that are not jointly orthogonal. Assume further that this set is minimal, in the sense that any proper subset is jointly orthogonal. Otherwise select a subset that is indeed minimal and redefine $n$. We may also assume that $n=3$. Otherwise redefine $A_3$ as the coarse-graining (disjunction) of $A_3,\ldots,A_n$. If the set $\{A_1,\ldots,A_n\}$ is a minimal non-Specker set, so is the new set $\{A_1,A_2,A_3\}$. 

Now take an ensemble of pairs of such systems in a maximally entangled state as specified in assumption (a). If all measurements performed are non-maximal measurements in the sense of assumption (b), we shall construct a protocol in which Alice can signal to Bob, thus violating (c). Note that the construction will go through even in the case $p_1+p_2+p_3\leq 1$ (otherwise we would merely have a proof of the exclusivity principle). The reason is that it does not rely on the incompatibility of $A_1,A_2,A_3$ in the sense that no joint probability measure for these three quantities exists, but in the sense that the joint probability for two of the propositions is generally disturbed by a measurement of the third.\footnote{Since $A_1,A_2,A_3$ are pairwise orthogonal, they must fail to be jointly orthogonal by failing to be jointly compatible. Because of (b) this means that -- although non-maximal measurements of each of $A_1,A_2,A_3$ exist  (again because of (b)), and although these can be performed in any order (because experiments can be performed sequentially) -- such a sequential measurement in general must fail to be a joint measurement of $A_1,A_2,A_3$. This in turn can only be the case if measuring one of them disturbs the joint probabilities for the other two.}

In our version of the fable Sandu and Daniel share pairs of maximally entangled systems with a three-element non-Specker set. Their propositions, however, are not only pairwise exclusive but also pairwise exhaustive. In this special case Sandu adopts the following protocol. If Daniel chooses $A$ and $B$ for his prophecy, Sandu chooses the \emph{third} box $C$, checks whether it is full or empty, and \emph{depending on the result} (regardless of whether he correctly prophesies it) he chooses which box to measure next.  If his $C$ is full, his next box will be empty, which implies that the corresponding box on Daniel's side will also be. And if Sandu's box $C$ is empty, his next box will be full, as will the corresponding one on Daniel's side. In this way, Sandu can ensure that Daniel's prophecy is always fulfilled. 

This protocol of course violates no-signalling (c), but we cannot use it in general because in our case pairs of propositions on the same side need not be exhaustive. Nevertheless, the protocol we now construct generalises the one in the fable: Alice measures one proposition, then depending on the result chooses to combine this with a measurement of one of the other two propositions she can measure. The freedom to do so presupposes condition (b). Given also the perfect correlations of condition (a), Alice will thereby affect the probabilities for a measurement of those other two propositions on Bob's side, thus violating condition (c). 

Let us proceed. Any two propositions $A_i$ and $A_j$ ($i=1,2,3$) can be measured together. In the maximally entangled state, the probability $p_i:=p(A_i=1)$ is strictly positive (because of (a)), and $p_i+p_j\leq 1$ (because the propositions are pairwise orthogonal). Unlike in the fable, there is no requirement that $p_i+p_j=1$: there may be some third proposition $A_k^*$ outside of the given set such that $A_i$, $A_j$ and $A_k^*$ are jointly orthogonal.

If a joint measurement of $A_i$ and $A_j$ is performed by either Alice or Bob, the conditional probabilities for the results are clearly:
  \be
    p(A_i=1|A_j=1)=0 \quad\mbox{and}\quad p(A_i=0|A_j=1)=1
    \label{proof1}
  \ee
(because the propositions are pairwise orthogonal) and
  \be
    p(A_i=1|A_j=0)=\frac{p(A_i=1\wedge A_j=0)}{p(A_j=0)}=\frac{p(A_i=1)}{p(A_j=0)}=\frac{p_i}{1-p_j}
    \label{proof2}
  \ee
(again because the propositions are pairwise orthogonal), from which we have also
  \be
    p(A_i=0|A_j=0)=1-\frac{p_i}{1-p_j}=\frac{1-p_i-p_j}{1-p_j} \ .
    \label{proof3}
  \ee
We may assume that the conditional probabilities on Bob's side are independent of whether or not a measurement of $A_k$ is performed by Alice. Otherwise, we already have a protocol by which Alice can signal to Bob, and there is nothing further to prove. Even so, if Alice measures $A_k$, the conditional probabilities on Bob's side need only remain the same on average, and may well depend on the outcome of Alice's measurement. We thus define:
  \begin{itemize}
    \item[(*)] Let $\alpha_{ij}^k$ and $\beta_{ij}^k\in[0,1]$ be the values of the conditional probability \mbox{$p(A_i=1|A_j=0)$} on Bob's side for the case in which Alice obtains outcome $A_k=1$ or $A_k=0$, respectively.
    \end{itemize}
No-signalling (c) imposes only the following restriction on $\alpha_{ij}^k$ and $\beta_{ij}^k$:\footnote{Note that we may assume $\alpha_{ij}^k$ and $\beta_{ij}^k$ are independent of whether or not Alice chooses to measure also $A_i$ or $A_j$ as well as measuring $A_k$: if they did depend on such a choice, then Alice could vary them separately by making different choices depending on the outcome of her measurement of $A_k$, thus violating (\ref{3*}) and thereby already violating (c).}
  \be
     p_k\alpha_{ij}^k+(1-p_k)\beta_{ij}^k=\frac{p_i}{1-p_j}\ .
     \label{3*}
  \ee   
     
We shall now show that even imposing (\ref{3*}) and whatever the values of $\alpha_{ij}^k$ and $\beta_{ij}^k$, Alice is able to signal to Bob.      
     
Assume that Alice attempts to affect the probabilities $p(A_1=1)$ and $p(A_2=1)$ on Bob's side by first measuring $A_3$, then choosing to measure either $A_1$ or $A_2$. Recall that $0<p_3<1$, so there are four possible cases:
  \begin{itemize}
    \item[(I)\ \ ] Alice obtains outcome $A_3$, then measures $A_1$;
    \item[(II)\ ] Alice obtains outcome $A_3$, then measures $A_2$;
    \item[(III)] Alice obtains outcome $\neg A_3$, then measures $A_1$;
    \item[(IV)] Alice obtains outcome $\neg A_3$, then measures $A_2$.
  \end{itemize}

In these four cases, the probabilities for $A_1=1$ and $A_2=1$ on Bob's side are, respectively:
  \begin{subequations}\label{proof4}
    \begin{align}
      &p^{\rm I}(A_1=1) =0 \ ,\mbox{ by (\ref{proof1}) on Alice's side and (a)},\\ 
      &p^{\rm I}(A_2=1) =\alpha_{21}^3\ ,\mbox{ by (*)};\\
      &p^{\rm II}(A_1=1) =\alpha_{12}^3\ ,\mbox{ by (*)},\\ 
      &p^{\rm II}(A_2=1)=0\ ,\mbox{ by (\ref{proof1}) on Alice's side and (a)};\\
      &p^{\rm III}(A_1=1)=\tfrac{p_1}{1-p_3}\ ,\mbox{ by (\ref{proof2}) on Alice's side and (a)},\\
      &p^{\rm III}(A_2=1)=\beta_{21}^3\cdot\tfrac{1-p_1-p_3}{1-p_3}\ ,\mbox{  by (\ref{proof3}) on Alice's side, (a) and (*)};\\
      &p^{\rm IV}(A_1=1)=\beta_{12}^3\cdot\tfrac{1-p_2-p_3}{1-p_3}\ ,\mbox{ by (\ref{proof3}) on Alice's side, (a) and (*)},\\ 
      &p^{\rm IV}(A_2=1)=\tfrac{p_2}{1-p_3}\ ,\mbox{ by (\ref{proof2}) on Alice's side and (a)}.
    \end{align}
  \end{subequations}

Assume more specifically that Alice wishes to minimise the value of \mbox{$p(A_1=1)$} on Bob's side. Using (\ref{proof4}) we now see that if she obtains $A_3=1$, choosing (I) over (II) gives the smaller probability because $0\leq\alpha_{12}^3$, indeed strictly so unless \mbox{$\alpha_{12}^3=0$}. Further, if instead she obtains $A_3=0$ it would be clearly useless for her to choose (III) over (IV), because then the total probability on Bob's side would be \mbox{$0\cdot p_3+\frac{p_1}{1-p_3}\cdot(1-p_3)=p_1$,} as if she had not performed any measurement at all. Thus, in order to have a chance of success, she has to choose (IV) over (III). And since she wishes to minimise $p(A_1=1)$, the worst-case scenario is when $\beta_{12}^3$ is maximal, i.e.\ again (by (\ref{3*})) when $\alpha_{12}^3=0$. 

In order to see whether Alice succeeds, we thus only need to check whether $\beta_{12}^3\cdot\frac{1-p_2-p_3}{1-p_3}<\frac{p_1}{1-p_3}$, or 
  \be
     \beta_{12}^3(1-p_2-p_3)<p_1
     \label{proof5}
  \ee 
for the worst-case scenario $\alpha_{12}^3=0$. By (\ref{3*}) this is the case when
  \be
    \beta_{12}^3=\frac{p_1}{(1-p_2)(1-p_3)}\ . 
    \label{proof6}  
  \ee
But inserting (\ref{proof6}) into (\ref{proof5}) it is immediately clear that
  \be
     \frac{p_1(1-p_2-p_3)}{(1-p_2)(1-p_3)}=\frac{p_1(1-p_2-p_3)}{1-p_2-p_3+p_2p_3}<p_1 
  \ee
since $p_2,p_3>0$ (again by assumption (a)), and Alice has succeeded in lowering the value of $p(A_1=1)$ on Bob's side.

Given the existence of a non-Specker set and assumptions (a) and (b), we have constructed a protocol by which Alice can signal to Bob, thus the assumption of a non-Specker set of propositions is inconsistent with the conjunction of assumptions (a), (b) and (c). \hfill\emph{QED}.

\section{The assumptions}\label{comments}
\subsection{`Maximal entanglement'}\label{a}
Our assumption (a) is meant to generalise the existence of maximally entangled states in quantum mechanics: we have perfect correlations between any pair of matching quantities (all of which have non-vanishing probability), even though the quantities on each side are not all jointly compatible.\footnote{This need not capture \emph{the} essence of maximal entanglement in quantum mechanics (see below in this subsection for an alternative generalisation), and a less loaded name for (a) could be `existence of perfect correlations', as suggested by another anonymous referee. On the other hand, the assumption is definitely non-classical because the propositions of each subsystem will generally not be jointly compatible. In this sense, the perfect correlations required by (a) are just as mysterious as the ones required by maximal entanglement in quantum mechanics.} 

In the case in which we have two sets of `Specker boxes' and  the same two boxes are opened on each side, say $A$ and $B$, Sandu and Daniel will either both find $A$ full and $B$ empty or they will both find $A$ empty and $B$ full. In the case in which only one of the boxes is the same on the two sides, say Sandu opens $C$ and $A$ and Daniel opens $C$ and $B$, then either both find $C$ full or both find $C$ empty.

In this second case, assumption (a) on its own does not yet determine what Sandu will find in $A$ and Daniel will find in $B$, but unless measurements on one side disturb the conditional probabilities on the other (thus violating (c)), it must be that if they both find $C$ full, they must both find their other box empty, and if they find $C$ empty they must find their other box full. Since $C$ and $A$ are compatible and $C$ and $B$ are compatible, if we assume also that opening individual boxes can be combined to yield a measurement of a pair of boxes (assumption (b)), then this will hold irrespective of the order in which boxes are opened. If Sandu opens first $A$ and then $C$ (e.g. finding $A$ empty and $C$ full), and Daniel opens first $B$ then $C$, Daniel will still find $B$ empty, because he still has to find $C$ full. And if first Sandu opens $A$ (e.g. finding it empty), then Daniel opens $B$, and then both open $C$, Daniel will also find $B$ empty, because they both have to find $C$ full.\footnote{If this scenario seems impossible because the choice of opening $C$ may still lie in the future, recall that the pattern of outcomes already seems impossible if the boxes are opened simultaneously. We postpone a full analysis of the scenario of the fable until Section~\ref{c}.}

Given that the boxes form a non-Specker set, our proof shows that assuming (a) and (b), one cannot consistently uphold (c). But our scenario is not the only one in the literature featuring some kind of `entangled Specker boxes'. Liang, Spekkens and Wiseman (2011, Section IV) consider a two-party one-query scenario, with two sets of three boxes but only \emph{one} box opened on each side. If the same box is opened on both sides, the results are perfectly correlated; if different boxes are opened, opposite results are obtained.\footnote{If this behaviour looks familiar, cf.\ Section~\ref{PR}.} They also consider a one-party two-query scenario in which one opens two boxes in succession on a single side, keeping fixed that one box turns out to be full and one empty (their Section III). If we combine these two scenarios, we get a two-party two-query scenario like in our fable, but the behaviour of the boxes is different. For instance, if $C$ is measured on either or both sides and found empty,  then subsequent measurements of $A$ or $B$ on either side will reveal them to be full (like in our fable). But if, for instance, $A$ and $B$ are measured on the two sides, in which case by assumption they produce opposite results, then subsequent measurements of $C$ on the two sides \emph{also} need to produce opposite results (\emph{unlike} in our fable). 

In terms of our three assumptions, it is clear that this new scenario \emph{satisfies no-signalling} (c): the average behaviour on each side is simply that of a single set of Specker boxes, irrespective of what measurements are made on the other side. Assumption (b) is also satisfied, because the order in which one opens two individual boxes does not affect the distribution of outcomes for the pair. What is violated (and has to be in order for Specker's principle to be violated) is our assumption (a): when $A$ and $B$ are measured on the two sides (yielding opposite results), the results of subsequent measurements of $C$ on the two sides are \emph{not} perfectly correlated, even though measuring $BC$ on one side and $CA$ on the other are both measurements of $C$. 

As suggested to me by Allen Stairs,\footnote{Email communication, 10 June 2017.} we can see this extended Liang, Spekkens and Wiseman scenario as 
exemplifying a different generalisation of maximal entanglement. Instead of requiring that measurements of the same quantity always give the same results on both systems, we can alternatively require that performing (compatible) measurements on one system affects the other system \emph{as if} the measurements had been performed on that system itself.\footnote{It might be interesting to investigate theories in which the two alternative versions of assumption (a) coincide.}

This last behaviour is somewhat analogous to that in the standard collapse formulation of quantum mechanics (where Specker's principle is of course satisfied). One can imagine that opening the first box (say, $A$ on Alice's side with outcome `empty') instantaneously `collapses' the joint state of all the boxes to a `disentangled' state (which we can write as $\ket{\mbox{empty},\mbox{full},\mbox{full}}\otimes\ket{\mbox{empty},\mbox{full},\mbox{full}}$). One can then take the resulting state on each side as locally determining the result of the next measurement. This will generally induce some further collapse, but again only locally. For instance, a measurement of $B$ on Bob's side will have outcome `full' with certainty and collapse the state to $\ket{\mbox{empty},\mbox{full},\mbox{full}}\otimes\ket{\mbox{empty},\mbox{full},\mbox{empty}}$, so that further measurements of $C$ on the two sides will have indeed outcomes `full' and `empty', respectively.\footnote{
One should not take the analogy too far. The toy model is formally satisfactory only if the maximally entangled state is its only state. If we extend it to include also collapsed states such as $\ket{\mbox{empty},\mbox{full},\mbox{full}}\otimes\ket{\mbox{empty},\mbox{full},\mbox{full}}$, then $B$ and $C$ will no longer be compatible, because measuring them on the same side produces different outcomes depending on the order of the measurements. For the closest quantum analogue, think of measurements of spin along three directions pairwise at angle $\alpha$ on a pair of spin-$1/2$ systems in the singlet state. These are pairwise incompatible observables, but their commutator is zero in the singlet state. (Instead, the quantum analogue for the scenario in our fable is given by measurements on a pair of spin-1 systems in a maximally entangled state, where perfect correlations or anticorrelations between corresponding one-dimensional projections are obtained independently of whatever other compatible measurements are performed alongside -- see also Section~\ref{b}.)}

\subsection{`Non-maximal measurements'}\label{b}
Assumption (b) is a generalisation of the existence of non-maximal measurements in quantum mechanics. A well-known example of the latter is the one used by Kochen and Specker (1967) to illustrate their famous theorem. Take any three orthogonal directions $x,y,z$. The projections $\ket{0_x}\bra{0_x},\ket{0_y}\bra{0_y},\ket{0_z}\bra{0_z}$ onto the eigenstates of zero spin in the three directions form a resolution of the identity, i.e.\ a quantum observable, in the Hilbert space of a spin-1 system. One way of measuring this observable is to combine in any order three separate measurements of the squared spin operators $S^2_x$, $S^2_y$, $S^2_z$, which are implemented by the Kraus operators $\ket{0_x}\bra{0_x}$ and $\iden-\ket{0_x}\bra{0_x}$ (and similarly for $y$ and $z$).
Note that the measurement of $S^2_x$ is combinable with the analogous measurements of $S^2_{y'}$ and $S^2_{z'}$ for \emph{any} other directions $y',z'$  that form an orthogonal triple with $x$. No other measurement of $\ket{0_x}\bra{0_x}$ is combinable with them, because any other resolution of the identity containing $\ket{0_x}\bra{0_x}$ will contain three mutually orthogonal projections and thus in general be \emph{in}compatible with measurements of other squared spin operators $S^2_{y'},S^2_{z'}$. In this sense, these measurements are measurements of $\ket{0_x}\bra{0_x}$ alone and thus indeed `non-maximal'. Note that they  exist for any projections in Hilbert space, but not for effects.\footnote{The existence of such measurements in quantum mechanics appears obvious today, but when von Neumann (1927) originally formalised quantum mechanics in Hilbert space (see Bacciagaluppi 2022), he assumed that a measurement of a multidimensional projection was always implemented through measurements of its maximal fine-grainings (subject to a non-contextuality condition). The modern notion of non-maximal measurements was introduced only by L\"uders (1950).}  

We shall now see what kind of violation of Specker's principle we can construct by dropping assumption (b). 

Note first of all that for Sandu to be able to signal to Daniel, it is crucial that condition (b) allows him to choose to open $A$ or $B$ \emph{after} having opened $C$ and seen the result. The reason is that, because of (c), the probabilities for Daniel's outcomes are in fact independent of \emph{whether or not} Sandu performs a measurement of $BC$ or of $CA$ (i.e.\ averaging over the outcomes of Sandu's measurement). It is only the \emph{conditional} probabilities for Daniel's outcomes given Sandu's that depend on Sandu's choice.\footnote{This dependence can be phrased 
as a dependence on 
the local measurement context on Sandu's side, by regarding his procedures as measurements of $C$ in the context of also measuring $B$ or also measuring $A$.  (We shall discuss this further in Section~\ref{c}.)} The way the story is told, Sandu's ability to make this choice appears trivial. That it is not so can be seen by adapting what is usually taken to be a different version of the same example, namely the `firefly box'.\footnote{My very special thanks to Alex Wilce for pressing me on this point, and on the need to clarify the precise sense in which the firefly box violates (b). I also owe him the information that the firefly box was apparently invented by Dave Foulis one day that Eugene Wigner was visiting his research group, as an illustration of the kind of work they were doing.} 

Take a triangular box, with translucent sides, and consider the following three experiments. It is dark, and you approach the box from any of the three sides, holding a lantern in your hand: under these conditions you always observe that something starts glowing in either of the two corners you can observe (in each case with probability $1/2$). This is the same example as Specker's three boxes in the sense that observations of any two corners are compatible with each other, the results of these observations are always opposite, and of course there is no joint probability distribution with the given marginals. But the way the story is told in the case of the firefly box, $C$ can only be measured \emph{simultaneously} with $A$ or with $B$. 

The name `firefly box' comes about because there is a simple mechanism for explaining the results of your observations. You imagine that somewhere in the box there sits a firefly, which mistakes the glow of your lantern for a potential mating partner, so \emph{moves all the way} to the side of the box you are approaching from and starts glowing! Thus, while each experiment corresponds in fact to a random variable on the space of initial positions of the firefly, the three experiments correspond to \emph{three different random variables}. In particular, the event `$C$ glowing when we approach from $CA$' is different from the event `$C$ glowing when we approach from $BC$'. This means that even though there is no non-contextual model for the three experiments, the firefly mechanism provides a \emph{contextual hidden variables model} for the experiments. The model is rather like pilot-wave theory, in which particle position is the hidden variable determining contextually the results of all experiments. 

Why does this model violate (b)? First of all, as already pointed out, there are no measurements of $C$ that are not also measurements of $A$ or of $B$ (and similarly for measurements of $A$ and of $B$). But further -- analogously to the quantum case -- a measurement of $C$ that is also a measurement of, say, $A$ will \emph{disturb} $B$. This is easy to see, because after a measurement of $CA$ the firefly will be somewhere along the $CA$-side of the box. But that means that, whether we perform $BC$ or $AB$, corner $B$ will \emph{always} remain dark. Therefore, performing a measurement of $C$ in the context of measuring also $A$ followed by a measurement of $B$ in the context of measuring either $C$ or $A$ does not yield a joint measurement of $B$ and $C$, and the measurement of $CA$ fails to qualify as a `non-maximal measurement' of $C$ in the sense of (b). Similarly, if we choose to measure $C$ in the context of measuring also $B$ (instead of $A$), then $A$ will be disturbed, and also $BC$ fails to be a non-maximal measurement of $C$ in the sense of (b). Since $CA$ and $BC$ are the only two measurements of $C$, the set ${\cal C}_C$ is empty, and similarly the sets ${\cal C}_A$ and ${\cal C}_B$, so the firefly box does not allow any non-maximal measurements at all in the sense of (b).

Now we can construct a bipartite system of two Specker-violating firefly boxes satisfying both (a) and (c). First of all, we can impose that the two firefly boxes be perfectly correlated in the sense of our assumption (a): say, if corner $C$ is observed both on Alice's firefly box (say together with $B$) and on Bob's firefly box (say together with $A$), then $C$ will glow on Alice's side if and only if it glows on Bob's side. Note that implementing this via the motion of the fireflies will require action at a distance. (Again like in pilot-wave theory!)\footnote{Specifically, once the first of the two entangled fireflies moves to a certain position along the sides of its box (because of a measurement), the other one must also move to the same position along the sides of its own box, then `cut the nearest corner'. E.g.\ if Alice measures $CA$, her firefly moves either to the $C$-half or the $A$-half of the $CA$-side of her box, and this forces Bob's firefly to move either to the $C$-half of the \emph{$BC$-side} of his box, or the $A$-half of the \emph{$AB$-side} of his box.}    

This is analogous to the case of our fable, but since the only procedures that Alice can implement are simultaneous observations of two corners, Alice cannot exploit the context-dependence of the conditional probabilities in order to signal to Bob, and condition (c) is also satisfied (one can also check explicitly that the resulting probabilities are non-signalling\footnote{Cf.\ again Section~\ref{PR}.}). Again as in pilot-wave theory, there is action at a distance, but (as long as Alice and Bob remain ignorant of the initial position of their fireflies!) this action at a distance \emph{cannot be used for signalling}.\footnote{For an insightful description of the aspects of pilot-wave theory that form the basis for the analogy, see Barrett (1999, Chapter~5). See in particular his description of measurements of spin, where the hidden variable is the initial position of the particle, the results of measurements are context-dependent (since they depend on the choice of polarity of the magnetic field, or of direction of the field gradient), the contextuality yields action at a distance in the case of entangled pairs, and we are unable to signal if we have no knowledge of the initial positions.} 

Finally, this analysis of the firefly box might also suggest a possible interpretation for assumption (b). Indeed, what is happening here is that we originally misidentify, say, `$C$ glowing when we approach from $CA$' and `$C$ glowing when we approach from $BC$' as one and the same event `$C$ glowing' because they are equiprobable. However, once we consider the hidden mechanism behind the experiments, we recognise that they are two different events (which can be distinguished given a non-standard distribution for the hidden variables). That is, even though there might initially appear to be a well-defined quantity $C$ because probabilities for outcomes of measurements of $C$ are independent of the chosen measurement procedure ($BC$ or $CA$), there is in fact \emph{no} such physical quantity independent of the context of observation. In the firefly example, thus, the failure of assumption (b) (the fact that there exist no measurements of $C$ alone even though $C$ is thought to be a well-defined physical quantity) indicates that $C$ is not after all a genuine physical quantity. 
One may thus reasonably expect that (b) should be a \emph{necessary condition} for an observable to be a \emph{genuine physical quantity}.\footnote{That it is not a \emph{sufficient} condition can be seen by noticing that we \emph{can} modify the entangled firefly boxes so that they satisfy (b) after all. What we need is to require that \emph{Alice's} firefly `cuts the nearest corner' after moving to the observed side (the motion we described in footnote~21 
for Bob's firefly). In that case the choice of ${\cal C}_A=\{AB\}$,  ${\cal C}_B=\{BC\}$ and ${\cal C}_C=\{CA\}$ or alternatively (!) the choice of ${\cal C}_A=\{CA\}$,  ${\cal C}_B=\{AB\}$ and ${\cal C}_C=\{BC\}$ will satisfy (b) (even though `non-maximal' may be a misnomer in this case). Depending on the behaviour of Bob's firefly, we can now violate (a) and satisfy (c), or satisfy (a) and violate (c). Specifically, if Bob's firefly moves as described in footnote~21 
upon the first of Alice's measurements, then moves again only when Bob performs any further local measurements, then we reproduce the behaviour of our extended Liang, Spekkens and Wiseman scenario, which violates (a) and satisfies (c). If instead Bob's firefly continues to mirror the motion of Alice's firefly also when Alice performs her second measurement, then Alice can choose to influence at a distance the behaviour of Bob's firefly depending on the result of her first measurement, and we reproduce the behaviour in our fable, which satisfies (a) and violates (c).}

\subsection{No-signalling}\label{c}
We finally return to assumption (c) and our fable of the seer of Nineveh. The condition of no-signalling is well known, but in the context of our fable its violation is nevertheless unusual. This is because, as in Specker's original version, the mechanism that suggests itself to explain the results is not a nonlocal mechanism such as instantaneous collapse or action at a distance, but \emph{retrocausation}.\footnote{It need not be, however: as noted in the previous footnote, the same behaviour as in our fable can be achieved also with action at a distance.} As devised by the father, the trials are meant to test whether the suitors are better prophets than himself (only then would he gladly surrender his daughters to them). But the father is an extremely good prophet indeed, and always prophesies correctly \emph{which two boxes will be opened on each table} (say, $AB$ on both sides, or $AB$ and $BC$), and fills the boxes accordingly. 
This means that opening the boxes the next day retrocausally influences which boxes are filled the previous night. 

Specifically, the father flips a coin to decide where to place the first  gem (say, in either $A$ or $B$ on the one side), and he places another gem accordingly on the other table (so as to satisfy (a)).\footnote{The model can be generalised further in various ways. Knowing which four boxes are to be opened on each trial allows the father to place gems according to any arbitrary distributions he might choose for any two pairs of boxes. Also, one can restrict or extend the model to the case in which fewer or more than two boxes get opened on at least one side. Clearly, based on the proportion of gems he has placed in, say, the $A$-boxes in the cases in which two boxes are opened on both sides, the father can place the same proportion of gems in the $A$-boxes for the case only these are opened on one side. And if we understand incompatibility in the sense that observed statistics depend on whether other measurements have been performed, it does not matter once two boxes have been opened on one side what the probability for finding the third one full is or what the correlations between the second and the third are, so the father can place further gems in the third box as he sees fit.} Assumption (b) is also clearly satisfied: opening two boxes is just the combination of opening the boxes singly, and whether a box is empty or full is a perfectly genuine physical question, albeit one that can be influenced retrocausally. And now, if Daniel makes a prophecy about $A$ and $B$, once Sandu knows that his $C$ is, say, empty, he knows that by choosing his second box to be, say, $B$, he will be causing the father to have put a gem inside it the previous night -- as well as a gem in the corresponding box on Daniel's side (and similarly in the other cases). This is how he manages to freely choose to make Daniel's prophecy turn out to be correct, thereby violating (c). Signalling is not across from one set of boxes to another, but in a zig-zag from Sandu's opening of his box back to the father's filling of the boxes and forward again to Daniel's opening of his.\footnote{Huw Price is a notable champion of retrocausation, and his Price (1996) provides a lucid discussion of how retrocausation makes sense in the first place and of how it might provide an explanation for the violations of the Bell inequalities in quantum mechanics. It should be noted that, while the fable of the seer provides an excellent illustration of some of the implications of retrocausation, it remains silent on the further mechanism behind the father's prophetic gift. For all we know, the causal link from opening the boxes one day and filling them the previous night might be operating at temporal distance. If that were a generic feature of retrocausal mechanisms, why prefer them to action at \emph{spatial} distance in order to explain nonlocal quantum correlations? But that is just an unhelpful association evoked by the idea of prophecy (perhaps a bit less if we call it `second sight'). The reason one might expect retrocausal effects in nature lies in the time symmetry (or CPT symmetry) of the laws of physics: for any process that may be given a description in terms of forwards causation, there is a corresponding process that can be given a description in terms of backwards causation. Thus one can conceive of an agent who may be able to exploit such backwards causal processes to affect events in their own past. If that is so, however, then retrocausal mechanisms to be expected on the basis of such time-symmetry arguments will be \emph{just as local} as all familiar forwards causal mechanisms. If they are at all nonlocal in time it will be on the basis of physical processes that are equally nonlocal in the forwards time direction, e.g.\ non-Markovian stochastic evolution. (Many thanks to Wayne Myrvold for raising this issue and to Huw Price for discussion.)} 

Relatedly, Sandu's ability to signal is not due to a violation of the well-known condition of parameter independence: \emph{given} the complete state of the boxes (i.e.\ whether they are full or empty), the probability for Daniel finding a gem in a particular box does not depend on anything that Sandu might do. In terms of the probabilistic conditions standardly used in discussing distant correlations in quantum mechanics, our version of the fable involves neither violations of outcome independence (as in collapse theories or in our version of the Liang, Spekkens and Wiseman scenario) nor violations of parameter independence (as in pilot-wave theory or in our entangled firefly boxes).  It rather involves violations of \emph{measurement independence}: the distribution of gems in the boxes depends on the choice of the future measurement contexts, and it is this mechanism that mediates the dependence of Daniel's outcomes on Sandu's choice of procedure.\footnote{For a detailed treatment of the relation between measurement dependence and signalling in the context of hidden variables theories, see Bacciagaluppi, Hermens and Leegwater (in preparation).}

A further striking feature of the entangled Specker boxes in our fable is that -- unlike the original case of only one daughter and only one suitor at a time, in which the retrocausal mechanism remains hidden -- in the entangled version retrocausation becomes \emph{manifest}. Indeed, we have chosen a case in which it becomes exploitable for signalling, but one can imagine even weirder things happening. Consider Daniel opening box $A$ and getting, say, $+1$ and Sandu opening box $B$ and getting, say, $-1$. We have seen that, given any possible choices of two boxes on either side, the distribution of gems in the boxes is such as to ensure \emph{both} perfect anticorrelations on each side \emph{and} perfect correlations across the two sides. It follows that Sandu and Daniel \emph{cannot both open box C}, because their boxes would have to be both full and empty. If Sandu and Daniel wanted to thwart the father's prediction, they would not be able to. Something (presumably perfectly natural) would always prevent them from opening box $C$ -- just like something always prevents you from killing your grandfather if you travel back to a time before your parents' birth. But there is nothing mysterious about it: Daniel's box $A$ being full and Sandu's box $B$ being empty simply reflects that the father knew already the previous night that Sandu and Daniel in fact would \emph{not} both open box $C$.\footnote{This is also the reason why, somewhat mystifyingly, at the end of the original fable the third box cannot be opened: \emph{not} that the three boxes can never be opened simultaneously or in sequence (incompatibility merely means that opening two of them must disturb the statistics of the third), but that on this particular occasion the father had correctly prophesied that precisely \emph{those two} would be opened. The cause for the failure to prise open the third is quite separate and presumably mundane, like the lid being jammed or the lock being defective.}

Finally, as remarked already (in footnote~19
), Sandu's strategy relies on the fact that the conditional probabilities for Daniel's outcomes given Sandu's outcomes will depend on Sandu's local measurement context (regarding Sandu's procedure as a measurement of $C$ in the context of measuring also $A$ or of measuring also $B$). The unconditional probabilities for Daniel's outcomes are instead unaffected by Sandu performing these measurements (which are indeed measurements of $CA$ or of $BC$). But we can equivalently think of Sandu's procedure as a single \emph{generalised} measurement (analogous to a quantum mechanical POVM) that \emph{does} affect the unconditional probabilities on Daniel's side. Specifically, if $A$, $B$ and $C$ were some self-adjoint operators with spectral resolutions 
\begin{equation}
A=A_+-A_-\ ,\qquad B=B_+-B_-\qquad\mbox{and}\qquad C=C_+-C_- \ , 
\label{eq1}
\end{equation}
Sandu's procedure (say, if he wants to guarantee that Daniel's box $A$ is empty) would simply be a measurement of the POVM with resolution of the identity
\begin{equation}
C_+A_+C_+ + C_+A_-C_+ + C_-B_+C_- + C_-B_-C_- = \iden
\label{eq2}
\end{equation}
(and similarly with $C_+$ and $C_-$ interchanged if he wants to guarantee that Daniel's box $A$ is full). Note that although the protocol in the fable involves some classical communication (because Sandu wants to know whether he should perform (\ref{eq2}) or interchange the roles of $C_+$ and $C_-$), this is not essential to ensure signalling. Already the simple fact of performing one of these two generalised measurements affects the probabilities for the outcomes of a possible measurement of $A$ and $B$ on Daniel's side.  

\begin{center}***\end{center}

In conclusion, the three examples we have described in this section -- the extended Liang, Spekkens and Wiseman scenario, the entangled firefly boxes, and the tale of the two suitors -- can be seen as three canonical examples of violations of Specker's principle obtained by relaxing in turn the three assumptions (a), (b), (c) of our theorem, and provide further insights into the meaning of these assumptions.

\section{Popescu and Rohrlich meet Specker's seer}\label{PR}
The two suitors in our tale are named after Sandu Popescu and Daniel Rohrlich, who in their `Quantum nonlocality as an axiom' (Popescu and Rohrlich 1994) investigated the combination of nonlocality with no-signalling. As is well known, Popescu and Rohrlich discovered that nonlocality and no-signalling allow for correlations that are stronger than quantum mechanical ones, and their work inaugurated the field of investigating such super-quantum correlations and the ways of distinguishing them from quantum correlations. Their main example has come to be known as a `PR box'. One imagines a black box with two pairs of alternative settings $a, a'$ and $b,b'$ such that for each combination $ab, ab', a'b, a'b'$, the box produces a pair of random results $\pm 1$, each individual result having probability 1/2, and one pair being perfectly correlated and the other three perfectly anti-correlated (or \emph{vice versa}). This in fact yields eight different PR boxes, each of which saturates the $S=4$ bound in the Bell inequalities. 

The scenario by Liang, Spekkens and Wiseman (2011) is an example of a realisation of a PR box when one considers opening alternatively boxes $A$ and $B$ on one side and $A$ and $C$ on the other -- which yields perfect correlations for the pair $AA$ and perfect anticorrelations for the other three pairs. 

The  scenario in our fable also provides such a realisation (and the seer must have found it amusing to prophesy that another Sandu and another Daniel were going to give their names to these contraptions). Let Daniel open either $A$ and $B$, or $C$ and $A$ on his side (label these two choices $a$ and $a'$), and let Sandu open either $A$ and $B$, or $B$ and $C$ on his side (label these two choices $b$ and $b'$). If we interpret Daniel's measurements, respectively, as a measurement of $A$ (in the context of measuring also $B$) or of $C$ (in the context of measuring also $A$), and Sandu's measurements, respectively, as a measurement of $A$ (in the context of measuring also $B$) or of $B$ (in the context of measuring also $C$), then we get perfect correlations for the pair $ab$ and perfect anticorrelations for the other three pairs of measurements. Thus we have again a realisation of a PR box. (The argument applies equally to entangled firefly boxes, since we have not used assumption (b).)

In fact, we can realise all eight PR boxes, as suggested to me by Jeff Bub.\footnote{Private communication, Tarquinia (Italy), May 2017. The argument again applies equally to entangled firefly boxes.} Indeed, by  changing the interpretation of Daniel's $a$ to that of a measurement of $B$ (in the context of measuring also $A$), we get perfect correlations for the pair $ab'$ and perfect anticorrelations for the other three. If we further change the interpretation of Sandu's $b'$ to that of a measurement of $C$ (in the context of measuring also $B$), we get perfect correlations for the pair $a'b'$ and perfect anticorrelations for the other three. If we change back the interpretation of Daniel's $a$, we get perfect anticorrelations for the pair $a'b$ and perfect correlations for the other three. Continuing in this fashion by considering Daniel's and Sandu's measurements alternatively as measurements of the first or of the second box (in the context of  opening also the other one), we get realisations of all eight Popescu--Rohrlich boxes (each of them twice --  because reinterpreting both of Daniel's and Sandu's measurements in fact yields again the same box).


Popescu and Rohrlich's original aim was to explore whether the combination of nonlocality (here implemented by our (a)) with no-signalling (our (c)) could be used to characterise quantum mechanics. Our proof of Specker's principle shows that combining these axioms with a condition on what should count as a genuine physical quantity (our (b)) comes one modest step closer to realising this aim.

\section*{\normalsize Acknowledgements}\vspace{-4ex}
\footnotesize This paper has had a long gestation, originating on the one hand in the pedagogical use of Specker boxes, firefly boxes and related examples in my teaching, various talks, and my handbook article (Bacciagaluppi 2016); on the other in an unfinished plan to discuss the comparison between the uses of Specker's principle in the older and newer literature in joint work with Alex Wilce, to whom my special thanks go for extensive discussion and feedback -- especially on the clarification of the general framework I am employing and my specific assumption (b) -- and for being my habitual consultant on matters quantum logical. I am further grateful to Ehtibar Dzhafarov and Sonja Smets for organising two conferences where this material was presented in preliminary form (and to the very perceptive audiences), and thank Ehti also for the invitation to contribute this paper to the present special issue, as well as two very helpful anonymous referees. Finally, I am very much indebted to Jeff Bub, Ad\'{a}n Cabello, Philipp H\"{o}hn, Huw Price, Allen Stairs and Ken Wharton for discussion of ideas related to this paper and/or comments on previous versions.

\section*{\normalsize References}\vspace{-4ex}
\footnotesize
\noindent Bacciagaluppi, G. (2016), `Quantum probability: An introduction', in A. H\'{a}jek and C. Hitchcock (eds), \emph{The Oxford Handbook of Probability and Philosophy} (Oxford: Oxford University Press), pp.~545--572. Extended version at \url{http://philsci-archive.pitt.edu/10614/}.

\noindent Bacciagaluppi, G. (2022), `The statistical interpretation: Born, Heisenberg and von Neumann, 1926--27', in O. Freire \emph{et al}. (eds), \emph{The Oxford Handbook of the History of Quantum Interpretations} (Oxford: Oxford University Press), pp.~203--231.

\noindent Bacciagaluppi, G., Hermens, R., and Leegwater, G. (in preparation), `Measurement dependence and nonlocality'.

\noindent Barnum, H., M\"uller, M., and Ududec, C. (2014), `Higher-order interference and single-system postulates characterizing quantum theory', \emph{New Journal of Physics} {\bf 16}, 1230291--44.

\noindent Barrett, J. A. (1999), \emph{The Quantum Mechanics of Minds and Worlds} (Oxford: Oxford University Press).

\noindent Cabello, A. (2012), `Specker's fundamental principle of quantum mechanics', \url{https://arxiv.org/abs/1212.1756}.

\noindent Cabello, A. (2013a), `Twin inequality for fully contextual quantum correlations', 
 \emph{Physical Review A} {\bf 87}, 010104(R)/1--4.
 
\noindent Cabello, A. (2013b), `Simple explanation of the quantum violation of a fundamental inequality', \emph{Physical Review Letters} {\bf 110}(6), 060402/1--5.

\noindent Cabello, A. (2015), `Simple explanation of the quantum limits of genuine $n$-body nonlocality', \emph{Physical Review Letters} {\bf 114}, 220402/1--5.

\noindent Cabello, A. (2019), `Quantum correlations from simple assumptions',  \emph{Physical Review A} {\bf 100}, 032120/1--14.

\noindent Cabello, A., Severini, S., and Winter, A. (2014), `Graph-theoretic approach to quantum correlations', \emph{Physical Review Letters} {\bf 112}(4), 040401/1--5.

\noindent Chiribella, G., Cabello, A., Kleinmann, M., and M\"uller, M. (2020), `General Bayesian theories and the emergence of the exclusivity principle', \emph{Physical Review Research} {\bf 2}(4), 042001/1--5.

\noindent Chiribella, G. and Yuan, X. (2014), `Measurement sharpness cuts nonlocality and contextuality in every physical theory', \url{https://arxiv.org/abs/1404.3348}. 

\noindent Finch, P. D. (1969), `On the structure of quantum logic', \emph{Journal of Symbolic Logic} {\bf 34}, 275--282. Reprinted in Hooker (1975), pp.~415--425.

\noindent Foulis, D. J., and Bennett, M. K. (1994), `Effect algebras and unsharp quantum logics', \emph{Foundations of Physics} {\bf 24}, 1331--1352. 

\noindent Foulis, D. J., and Randall, C. H. (1981), `Empirical logic and tensor products', in H. Neumann (ed.), \emph{Interpretations and Foundations of Quantum Mechanics} (Mannheim: Bibliographisches Institut), pp.~9--20.

\noindent Fritz, T., Sainz, A. B., Augusiak, R., Bohr Brask, J., Chaves, R., Leverrier, A., and Ac\'{i}n, A. (2013), `Local orthogonality as a multipartite principle for quantum correlations', \emph{Nature Communications} {\bf 4}, 2263/1--7.

\noindent Gudder, S. P. (1972), `Partial algebraic structures associated with orthomodular posets', \emph{Pacific Journal of Mathematics} {\bf 41}, 717--730.

\noindent Hardegree, G. M., and Frazer, P. J. (1981), `Charting the labyrinth of quantum logics: A progress report', in E. Beltrametti and B. van Fraassen (eds), \emph{Current Issues in Quantum Logic} (New York: Plenum Press), pp.~53--76.

\noindent Henson, J. (2012), `Quantum contextuality from a simple principle?', \url{https://arxiv.org/abs/1210.5978}. 

\noindent H\"{o}hn, P. A. (2017), `Toolbox for reconstructing quantum theory from rules on information acquisition', \emph{Quantum} {\bf 1}, 38/1--78.

\noindent Hooker, C. A. (1975), {\em The Logico-Algebraic Approach to Quantum Mechanics}, Vol.~1 (Dordrecht: Reidel).

\noindent Hughes, R. I. G. (1989), \emph{The Structure and Interpretation of Quantum Mechanics} (Cambridge, Mass.: Harvard University Press).

\noindent Kochen, S., and Specker, E. P. (1965), `Logical structures arising in quantum theory', in L. Addison, L. Henkin and A. Tarski (eds), {\em The Theory of Models\/} (Amsterdam: North-Holland), pp.~177--189. Reprinted in Hooker (1975), pp.~263--276.

\noindent Kochen, S., and Specker, E. P. (1967), `The problem of hidden variables in
quantum mechanics', \emph{Journal of Mathematics and Mechanics} {\bf 17}, 59--88.
Reprinted in Hooker (1975), pp.~293--328.

\noindent  Liang, Y. C., Spekkens, R. W., and Wiseman, H. M. (2011), `Specker's parable of the overprotective seer: A road to contextuality, nonlocality and complementarity', \emph{Physics Reports} {\bf 506}(1--2), 1--39.

\noindent L\"uders, G. (1950), `\"{U}ber die Zustands\"{a}nderung durch den Me\ss proze\ss ', \emph{Annalen der Physik} \mbox{{\bf 443}(5--8),} 322--328.

\noindent  Mackey, G. (1963), \emph{The mathematical foundations of quantum mechanics} (New York: Benjamin).

\noindent Neumann, J. von (1927), `Wahrscheinlichkeitstheoretischer Aufbau der Quantenmechanik', \emph{Nach\-richten von der Gesellschaft der Wissenschaften zu G\"{o}ttingen. Mathematisch-Phy\-si\-ka\-li\-sche Klasse} \mbox{{\bf 1927}(11 November),} 245--272.

\noindent Popescu, S., and Rohrlich, D. (1994), `Quantum nonlocality as an axiom', \emph{Foundations of Physics} {\bf 24}(3), 379--385.

\noindent Price, H. (1996), \emph{Time's Arrow and Archimedes' Point: New Directions for the Physics of Time} (Oxford: Oxford University Press).

\noindent Seevinck, M. P. (2011), `The logic of non-simultaneously decidable propositions', \url{http://arxiv.org/abs/1103.4537}. Translation of Specker (1960).

\noindent Specker, E. P. (1960), `Die Logik nicht gleichzeitig entscheidbarer Aussagen', {\em Dialectica\/} {\bf 14}, 239--246. Translated as Stairs (1975) and as Seevinck (2011). 

\noindent Stairs, A. (1975), `The logic of propositions which are not simultaneously decidable', in Hooker (1975), pp~135--140. Translation of Specker (1960).

\noindent Wilce, A. (2009), `Test spaces', in D. Gabbay, D. Lehmann and K. Engesser (eds), \emph{Handbook of  Quantum Logic} (Amsterdam: Elsevier), pp.~443--549.

\end{document}